\documentclass[
pra,
superscriptaddress,
amsmath,amssymb,
aps,twocolumn
]{revtex4-2}

\usepackage{graphicx}
\usepackage{bm,amsfonts}
\usepackage[normalem]{ulem}

\usepackage{hyperref}
\DeclareMathOperator\arctanh{arctanh}
\usepackage[caption=false]{subfig}

\newcommand {\fabs}[1] {\left| #1 \right|}

\newcommand {\fabsq}[1] {\left\vert #1 \right\vert^{2}}

\newcommand{\ket}[1]{\ensuremath{|#1\rangle}}

\newcommand{\braket}[2]{\langle#1|#2\rangle}

\newcommand{\cE}{{\cal{E}}}
\newcommand{\cL}{{\cal{L}}}

\newcommand{\cF}{{\cal{F}}}

\newcommand{\Eqref}[1]{Eq. \eqref{#1}}

\newcommand{\ve}[1]{\boldsymbol{#1}}

\newcommand{\veuh}[1]{\mathbf{\hat{#1}}}

\newcommand{\asi}{a_{s}^0}
\newcommand{\asf}{a_{s}^f}

\newcommand{\wxi}{w_{x}^0}
\newcommand{\wxf}{w_{x}^f}
\newcommand{\wri}{w_{\rho}^0}
\newcommand{\wrf}{w_{\rho}^f}

\begin{document}


\title{Smooth time-dependent control of dipolar Bose-Einstein condensates}

\author{C. Whitty}
\affiliation{Department of Physical Chemistry, University of the Basque Country UPV/EHU, 48940 Leioa, Spain}
\email{c.whitty@ehu.eus}

\author{A. Ala\~na}
\affiliation{%
Department of Physics, University of the Basque Country UPV/EHU, 48080 Bilbao, Spain}

\author{M. Modugno}
\affiliation{%
Department of Physics, University of the Basque Country UPV/EHU, 48080 Bilbao, Spain}
\affiliation{EHU Quantum Center, University of the Basque Country UPV/EHU, 48940 Bilbao, Spain}
\affiliation{IKERBASQUE, Basque Foundation for Science, 48009 Bilbao, Spain}

\author{Xi Chen}
\affiliation{Quantum Advanced Research Center (QuARC), CSIC, 28049, Madrid, Spain}
\affiliation{Instituto de Ciencia de Materiales de Madrid (ICMM), CSIC, 28049, Madrid, Spain}

\author{G\'eza T\'oth}
\affiliation{Department of Physics, University of the Basque Country UPV/EHU, 48080 Bilbao, Spain}
\affiliation{EHU Quantum Center, University of the Basque Country UPV/EHU, 48940 Bilbao, Spain}
\affiliation{Donostia International Physics Center DIPC, Paseo Manuel de Lardizabal 4, 20018 San Sebastián, Spain}
\affiliation{IKERBASQUE, Basque Foundation for Science, 48009 Bilbao, Spain}
\affiliation{HUN-REN Wigner Research Centre for Physics, P.O. Box 49, 1525 Budapest, Hungary}

\author{A. Ruschhaupt}
\affiliation{School of Physics, University College Cork, Cork, Ireland}

\author{E. Ya. Sherman}
\affiliation{%
 Department of Physical Chemistry, University of the Basque Country UPV/EHU, 48940 Leioa, Spain}
\affiliation{IKERBASQUE, Basque Foundation for Science, 48009 Bilbao, Spain}
\affiliation{EHU Quantum Center, University of the Basque Country UPV/EHU, 48940 Bilbao, Spain}

\date{\today}

\begin{abstract}

We consider protocols for control of dipolar Bose-Einstein condensates where the critical role is played by the long-range anisotropic interatomic magnetic  dipole-dipole interaction.  The phase diagram of such a condensate has been explored  theoretically and experimentally with  certain values of the interatomic scattering length corresponding to superfluid and supersolid phases, where supersolidity appears as a modulation in the ground state density. Preparation of this modulated ground state is challenging, since excitations appear as a result of a finite-time evolution required to produce qualitative changes in the wavefunction density. To solve this problem we consider the time-dependent control of a dipolar Bose-Einstein condensate using shortcuts to adiabaticity techniques, concentrating on design of the time-dependent scattering length, a parameter of the system easily tunable by contemporary experiments. The first technique is the variational approach based on the Euler-Lagrange equations for a separable ansatz describing the evolution of the superfluid state. Secondly, we study the transition from superfluid to supersolid using a direct optimization protocol. We discuss the fidelity of the developed protocols in terms of the evolution time.  
\end{abstract}

\maketitle


\section{\label{sec:intro} Introduction}
Dipolar Bose-Einstein condensates (BEC) are a practical platform to investigate novel quantum many-body phenomena and phase transitions \cite{2009-PhysicsDipolarBosonic-Lahaye-Pfau,2012-CondensedMatterTheory-Baranov-Zoller,2020-NewStatesMatter-Bottcher-Pfau,2016-BoseEinsteinCondensation-Pitaevskij-Stringari,2022-DipolarPhysicsReview-Chomaz-Pfau, 2026-QuantumStabilizedStatesMagnetic-Chomaz}.
During the last two decades experimental control techniques have been developed for ultra-cold gases of several highly-magnetic atomic species, e.g. $^{166}$Er, $^{162}$Dy, $^{164}$Dy \cite{2022-DipolarPhysicsReview-Chomaz-Pfau, 2023-SupersolidityUltracoldDipolar-Recati-Stringari,2019-LongLivedTransientSupersolid-Chomaz-Ferlaino,2019-ObservationDipolarQuantum-Tanzi-Modugno,2019-TransientSupersolidProperties-Bottcher-Pfau}.
This rapidly improving experimental control has allowed for the observation of novel states of matter, in particular ultra-dilute droplets \cite{2016-SelfboundDropletsDilute-Schmitt-Pfau, 2016-ObservationQuantumDroplets-Ferrier-Barbut-Pfau,2024-MeasurementExcitationSpectrum-Houwman-Mark}, superfluid and supersolid phases \cite{2016-ObservingRosensweigInstability-Kadau-Pfau,2019-ObservationDipolarQuantum-Tanzi-Modugno,2019-LongLivedTransientSupersolid-Chomaz-Ferlaino,2019-TransientSupersolidProperties-Bottcher-Pfau,2019-SupersolidSymmetryBreaking-Tanzi-Stringari,2021-EvidenceSuperfluidityDipolar-Tanzi-Modugno,2021-BirthLifeDeath-Sohmen-Ferlaino,2023-HeatingDipolarQuantum-Sanchez-Baena-Pohl} and the superfluid-supersolid phase transition \cite{2022-DimensionalCrossoverSuperfluidSupersolid-Biagioni-Modugno, 2022-CrossingSuperfluid-supersolid-alana-modugno, 2024-SupersolidformationtimeShortcutExcitation-Alana-Alana}.
The finite-time transition between superfluid and supersolid phases produces excitations in the density distribution of the desired supersolid state, strongly dependent on the evolution protocols.
Thus, designing control schemes for the time-dependent evolution of dipolar Bose-Einstein condensates is a challenging task, and often quasi-adiabatic control schemes are used in practical settings \cite{2021-PhaseCoherenceOutofequilibrium-Ilzhofer-Ferlaino,2024-DipolarDropletsCrossover-Pylak-Zin,2021-BirthLifeDeath-Sohmen-Ferlaino}.

Time-dependent control of dipolar condensates has been previously considered, for example, quench dynamics in one and two dimensions \cite{2022-Control$^164mathrmDy$BoseEinstein-Halder-Sadeghpour,2023-ViscousDynamicsQuenched-Wang-Bohn,2023-SupersolidStacksAntidipolar-Mukherjee-Reimann}, control of one-dimensional droplets \cite{2018-DynamicsOnedimensionalQuantum-Astrakharchik-Malomeda,2025-QuantumDropletSpeed-Das-Nath}, and optimal control of the fast formation of self-bound dipolar droplets \cite{2015-OptimalControlBose-Mennemann-Langen,2019-OptimalControlSelfbound-Mennemann-Mauser,2022-SelfboundDipolarDroplets-Schmidt-Langen}.
A combination of bang-bang and quasi-adiabatic control performs well in the transition from superfluid to supersolid \cite{2024-SupersolidformationtimeShortcutExcitation-Alana-Alana}.
The quench dynamics of two-dimensional honeycomb and stripe phases has been explored \cite{2021-PhasesSupersolidsConfined-Zhang-Maucher,2025-DiracPointsShear-Blakie-Blakie}, where honeycomb crystals are shown to melt into several phases \cite{2025-DiracPointsShear-Blakie-Blakie}.

In line with recent experimental procedures, we consider the formation of a supersolid state as a controlled finite-time two-stage process, where the interatomic scattering length $a_s(t)$ is the time-dependent control parameter of the system \cite{2019-ObservationDipolarQuantum-Tanzi-Modugno,2019-LongLivedTransientSupersolid-Chomaz-Ferlaino,2019-TransientSupersolidProperties-Bottcher-Pfau,2019-SupersolidSymmetryBreaking-Tanzi-Stringari,2021-EvidenceSuperfluidityDipolar-Tanzi-Modugno}.
In the specific system considered here, there is a continuous phase transition from superfluid to supersolid about a critical value of the scattering length $a_s^c$ \cite{2019-ObservationDipolarQuantum-Tanzi-Modugno,2019-LongLivedTransientSupersolid-Chomaz-Ferlaino,2019-TransientSupersolidProperties-Bottcher-Pfau,2019-SupersolidSymmetryBreaking-Tanzi-Stringari,2021-EvidenceSuperfluidityDipolar-Tanzi-Modugno,2022-CrossingSuperfluid-supersolid-alana-modugno, 2024-SupersolidformationtimeShortcutExcitation-Alana-Alana}.
In the first stage, the system is initialized in a superfluid ground state for a scattering length value far from the superfluid-supersolid transition.
The system is then driven to a superfluid ground state close to the phase transition, using a derived time-dependent scattering length $a_{s}(t)$.
In the second stage, an $a_{s}(t)$ is designed to drive the system through the phase transition to a final value of the scattering $a_s < a_s^c$ in the supersolid regime.
Due to their significant differences, these stages require distinct control protocols, where the interplay between the non-linear interactions presents significant challenges.

In this paper we design control schemes for $a_{s}(t)$ using a Shortcut to Adiabaticity (STA) approach.  STAs are a collection of mainly analytic and semi-analytic techniques to control evolution of quantum systems.
They have been applied to a plethora of quantum systems \cite{2013-Chapter2Shortcuts-Torrontegui-Mugaa,2018-InverseEngineeringFast-Chen-Sherman,2019-ShortcutsAdiabaticityConcepts-Guery-Odelin-Muga}, and have been used in several BEC experiments \cite{2015-NonequilibriumScaleInvariance-Rohringer-Trupke, 2018-ShortcutLoadingBose-Zhou-Schmiedmayer}.
For the first stage we employ the STA variational approach with inverse engineering (VAIE) which has been applied to BECs without long-range interactions, including BEC expansion \cite{2020-ShortcutsAdiabaticityInteracting-Huang-Chen, 2021-EffectiveScalingApproach-Huang-Chen, 2025-ShortcutsAdiabaticityAnisotropic-Mishra-Fogarty}, fast transport \cite{2022-FastTransportBose-Li-Ruschhaupt}, control of soliton matter waves \cite{2016-ShortcutAdiabaticControl-Li-Chen}, and quantum heat engines \cite{2022-QuantumHeatEngine-Li-Ruschhaupt}.
The time-independent variational approach has been widely used to find the phase diagram of dipolar BECs in different realizations \cite{2012-NumericalVariationalSolutions-Muruganandam-Adhikari,2020-VariationalTheoryGround-Blakie-Pal,2021-NumericalCalculationDipolarquantumdroplet-Lee-Blakiea,2023-TwodimensionalSupersolidityPlanar-Ripley-Blakie,2023-SupersolidityCrystallizationDipolar-Smith-Blakie,2024-SoundWavesFluctuations-Platt-Blakie, 2015-GeneralTheoryFlattened-Baillie-Blakie}, and the time-dependent variational approach without inverse engineering has been applied to BECs with long-range interactions, for example, with non-local Gaussian and Van der Waals interactions \cite{2018-TimedependentVariationalApproach-Haas-Eliasson}, the dynamics of bright and dark solitons \cite{2016-DynamicsSolitonsOnedimensional-Ilg-Wunner, 2016-ExploringStabilityDynamics-Edmonds-Parker}, the dynamics of quantum droplets in Bose-Bose mixtures \cite{2024-DynamicsQuasionedimensionalQuantum-Otajonov-Abdullaev},the dynamics of dark solitons in a dipolar BEC \cite{2005-TwoDimensionalBrightSolitons-Pedri-Santos}, dipolar BEC collapse and the time-dependent energy spectrum \cite{2007-RadialAngularRotons-Ronen-Bohn, 2010-VariationalMethodsCoupled-Rau-Wunner1,2010-VariationalMethodsCoupled-Rau-Wunner2,2013-VariationalApproachBogoliubov-Kreibich-Wunner}, and the exploration of Faraday and resonant waves \cite{2019-FaradayResonantWaves-Vudragovic-Balaz}.
Although the above approaches explore the system's phase diagram and dynamics, they do not provide algorithms for desired control tasks.

In this paper we consider a time-dependent variational approach with inverse engineering, where one first assumes that the solution of the system's dynamics is well approximated by a time-dependent ansatz, a known analytical expression with a chosen set of parameters \cite{1997-VariationalThomasFermiTheory-Timmermans-Huang, 2002-EffectiveWaveEquations-Salasnich-Reatto}.
The governing equations of the system, in this case the nonlinear extended Gross-Pitaevskii equations (eGPE), are applied to the ansatz, and result in a system of simplified coupled equations that describe the system evolution approximately \cite{2020-VariationalTheoryGround-Blakie-Pal, 2020-SupersolidityElongatedDipolar-Blakie-Ferlaino}.
The boundary conditions imposed by the control task are then applied to these coupled equations.
From these equations the desired control can be inverse engineered by solving for the control parameter, in this case the time-dependent scattering length $a_s(t)$.
The combination of the variational approach with inverse engineering (VAIE) has the important advantage of being analytic by the choice of the ansatz, allowing significant freedom in the control scheme design, while minimizing intensive numerical computation, such as machine learning \cite{2025-MachineLearningCharges-King-Cheng}, or numerical pulse engineering \cite{2019-OptimalControlSelfbound-Mennemann-Mauser}.

The VAIE is efficient when designing control schemes in the superfluid regime, where the three-dimensional wavefunction density can be approximated well by a product of Gaussians.
However, for the transition between the superfluid and supersolid phases, this approach breaks down and an ansatz of greater complexity is required to pass between these two phases \cite{2025-KibbleZurekScalingSuperfluidsupersolid-Kirkby-Chomaza,2021-BraggScatteringUltracold-Petter-Ferlaino, 2026-QuantumStabilizedStatesMagnetic-Chomaz,2021-NumericalCalculationDipolarquantumdroplet-Lee-Blakiea}.
Since the goal of this work is to produce smooth and practical control schemes, we use a direct optimization approach for the superfluid to supersolid transition. Remarkably, a simple two-parameter numerical optimization of well-known polynomial adiabatic schemes can control the system with high fidelity.

The paper is organized as follows; in Section \ref{sec:model} we introduce the system, model, and the control task in detail.
In Section \ref{sec:GSSFSF} we apply the variational approach with inverse engineering to ground state transfer in the superfluid regime, and compare it with standard adiabatic control schemes.
The next Section presents optimization of the superfluid to supersolid transition.
We summarize our findings and give an outlook in the Conclusion.

\section{\label{sec:model} Model and Control Task}

\begin{figure}
\flushleft
\includegraphics[width=\linewidth]{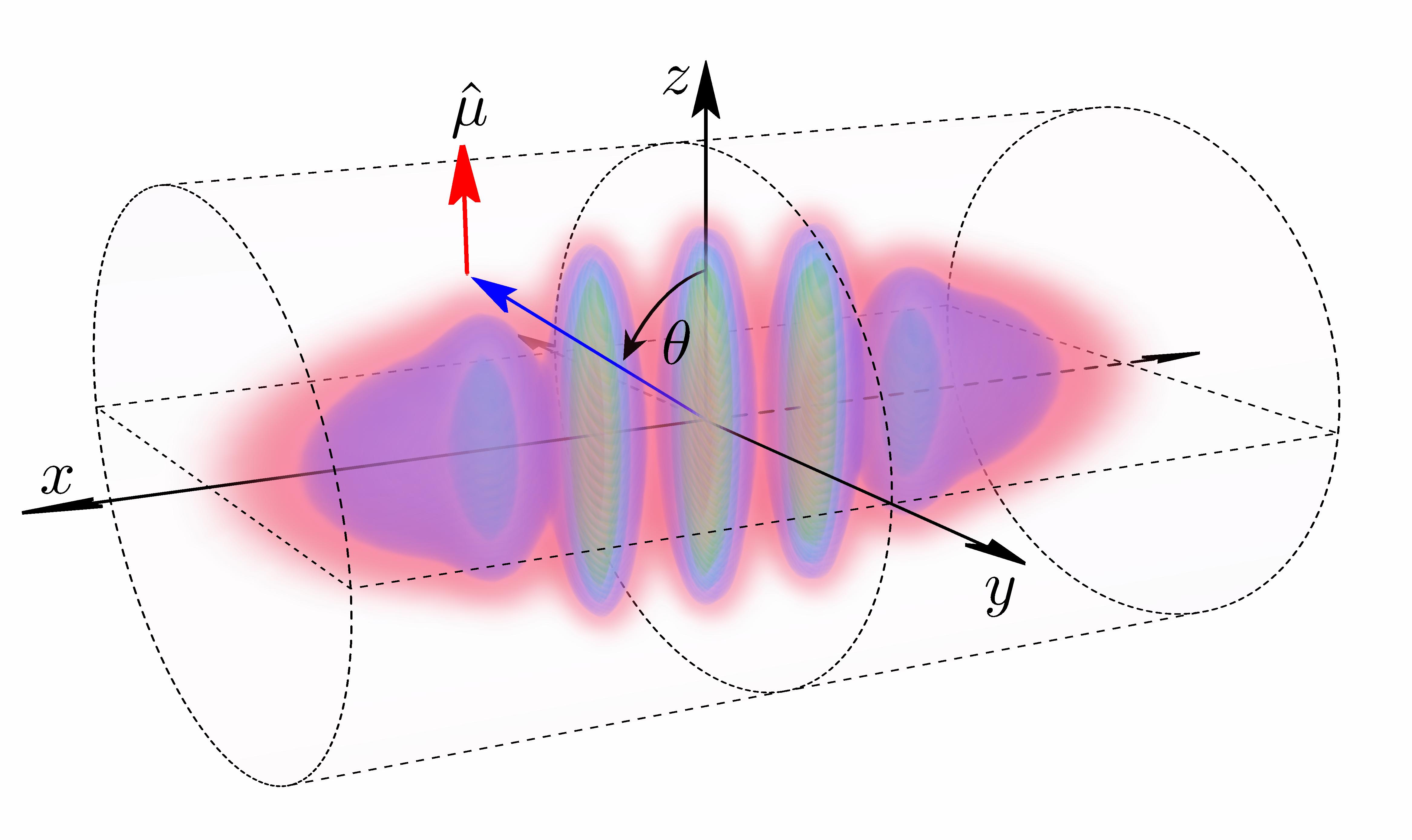}
\caption{Diagram of the system setup: The ground state density for the supersolid phase with $a_s=90.9 \; a_0$ is shown, with the cylindrical geometry of the system highlighted. The magnetic dipole moment is shown as $\veuh{\mu}$, oriented in the $\veuh{z}$ direction.}
\label{fig:system_setup.jpg}
\end{figure}
We consider $N$ bosons, each with mass $m$ described by the three-dimensional extended Gross-Pitaevskii equation \cite{2009-PhysicsDipolarBosonic-Lahaye-Pfau}
\begin{align}\label{eq:egpe}
i \hbar \frac{\partial}{\partial t}
&\Psi(\ve{r},t) =
\Big[
-\frac{\hbar^{2}}{2m}\nabla^{2}
+ V(\ve{r}) \nonumber \\
& + \frac{N \hbar^2}{m}\int d^3r' U_{dd} \left( \ve{r}-\ve{r}' \right) \fabsq{\Psi(\ve{r}',t)} \nonumber \\
& + g_s\left( t\right)  \left|\Psi(\ve{r},t)\right|^{2}
 + g_{L}\left( t\right) \left|\Psi(\ve{r},t)\right|^3
\Big] \Psi(\ve{r},t),
\end{align}
where $V(\ve{r}) = m(\omega_{x}^{2} x^{2} + \omega_y^{2} y^{2} +\omega_{z}^{2} z^{2})/2$ and $\int d^{3}r\fabsq{\Psi(\ve{r},t)}=1$.
The magnetic moments are aligned in the $\veuh{z}$ direction with the dipole-dipole interaction (DDI) potential given by \cite{2009-PhysicsDipolarBosonic-Lahaye-Pfau}
\begin{align}\label{eq:U_dip}
U_{dd} \left( \ve{r}\right) = \frac{C_{dd}}{4 \pi}\frac{1-3 \: \cos^{2}\theta}{\fabs{\ve{r}}^3},
\end{align}
where $C_{dd} = \mu_0 \mu^{2}$, $\mu$ is the magnetic moment of a single-atom and $\mu_0$ is the vacuum permeability.
A diagram of the the system's cylindrical geometry and the BEC ground state density in the supersolid regime, is shown in Fig. \ref{fig:system_setup.jpg}.
The contact interaction strength is
\begin{align}\label{eq:contact_interaction}
g_{s}(t) &= 4 \pi a_{s}(t)  N \hbar^{2} / m,
\end{align}
where $a_{s}(t)$ is the time-dependent $s$-wave scattering length.
A characteristic dipole length associated with the atomic species considered is given by $a_{dd}=(C_{dd} m)/(12 \pi \hbar^{2})$, and the ratio of $a_{dd}$ and the contact interaction length is $\epsilon_{dd}=a_{dd}/ a_{s}$, which characterizes whether the dipole interaction ($\epsilon_{dd}>1$) or the contact interaction ($\epsilon_{dd}<1$) dominates.

Corrections to the ground state energy of the BEC due to quantum fluctuations were first derived by Lee, Huang and Yang (LHY) \cite{1957-EigenvaluesEigenfunctionsBose-Lee-Yang, 1957-ManyBodyProblemQuantum-Lee-Yang}, and have been extended to the eGPE under the local density approximation \cite{2011-QuantumFluctuationsDipolar-Lima-Pelster, 2012-MeanfieldLowlyingExcitations-Lima-Pelster,1997-VariationalThomasFermiTheory-Timmermans-Huang,1997-QuantumCorrectionsGround-Braaten-Nieto,2015-TheoryExcitationsDipolar-Boudjemaa-Boudjemaa}, where the LHY interaction term is
\begin{align}
g_{L}\left( t\right) &= \frac{128 \sqrt{\pi} }{3}\frac{\hbar^{2}}{m}N^{3/2} {\:a_{s}^{5/2}(t)} \: Q_{5}(\epsilon_{dd}),
\end{align}
with \cite{2006-MeanFieldExpansionBose-Schutzhold-Fischer, 2011-QuantumFluctuationsDipolar-Lima-Pelster}
\begin{align}
Q_{5}(\epsilon_{dd}) = \frac{1}{2} \int_0^\pi d\theta \: \sin \theta
\left[ 
1 + \epsilon_{dd} \left( 3 \cos^{2} \theta - 1 \right)
\right]^{5/2}.
\end{align}
There are several approximations of $Q_5$ in the literature \cite{2016-QuantumFilamentsDipolar-Wachtler-Santos,2016-ObservationQuantumDroplets-Ferrier-Barbut-Pfau,2016-PathIntegralMonteCarlo-Saito-Saito,2016-GroundstatePhaseDiagram-Bisset-Blakiea}, and we use the simplified expression $Q_5(\epsilon_{dd}) \approx 1 + 3\epsilon_{dd}^{2}/2$ from \cite{2016-GroundstatePhaseDiagram-Bisset-Blakiea}, giving
\begin{align}\label{eq:LHY_interaction}
g_{L}\left( t\right) &= \frac{128 \sqrt{\pi }}{3} \frac{\hbar^{2}}{m}N^{3/2} a_{s}^{5/2}(t)\left(1 + \frac{3}{2}\frac{a_{dd}^{2}}{a_{s}^{2}(t)} \right).
\end{align}

We use dimensionless units based on a unit of length $l_0 = [\hbar / (m \omega_0)]^{1/2}$, where $\omega_0$ can be chosen for convenience.
We use a unit of length $l_0=1$ $\mu$m, with the corresponding harmonic oscillator angular frequency $\omega_0 = \hbar / (m l_0^{2})$ defining a unit of time $\tau_0= 1/\omega_0$, and unit of energy $\cE_0=\hbar \omega_0$.
We scale \Eqref{eq:egpe} using $\widetilde{t} = t / \tau_0$ and $\widetilde{\ve{r}} = \ve{r} / l_0$, with $\widetilde{\Psi}=l_0^{3/2} \Psi $.
The trap angular frequencies are scaled by $\omega_0$, i.e. $\widetilde{\omega}_j = \omega_j / \omega_0$ for $j=x,y,z$.
For notational convenience, we drop the tilde notation and use dimensionless units, noting the physical units explicitly where necessary.

A Lagrangian for \Eqref{eq:egpe} is given by $L\left[ \Psi,\Psi^*\right] = \int d^3r \: \cL(\ve{r},t)$, with the Lagrangian density
\begin{align}\label{eq:lag_dim}
\cL(\ve{r},t) &=  
\frac{i}{2} \left(
\Psi   \frac{\partial}{\partial t} \Psi^{*} - 
\Psi^* \frac{\partial}{\partial t} \Psi
\right) 
\nonumber \\
&+
\frac{1}{2} \fabsq{\nabla \Psi(\ve{r},t)} + V(\ve{r}) \fabsq{\Psi(\ve{r},t)}
\nonumber \\ 
&+\frac{N}{2} \int d^3r' U \left( \ve{r}-\ve{r}' \right) \fabsq{\Psi(\ve{r}',t)}
\fabsq{\Psi(\ve{r},t)}
\nonumber \\
& + \frac{1}{2} g_s\left( t\right) \fabs{\Psi(\ve{r},t)}^4
+ \frac{2}{5}  g_{L}\left( t\right) \fabs{\Psi(\ve{r},t)}^{5}.
\end{align}
This system undergoes a transition from a superfluid to supersolid phase, where a periodic modulation of density appears around a critical value of the $s$-wave scattering length $a_{s}^{c}$ \cite{2019-LongLivedTransientSupersolid-Chomaz-Ferlaino,2019-ObservationDipolarQuantum-Tanzi-Modugno,2019-SupersolidSymmetryBreaking-Tanzi-Stringari,2022-DimensionalCrossoverSuperfluidSupersolid-Biagioni-Modugno,2019-TransientSupersolidProperties-Bottcher-Pfau,2020-SupersolidityElongatedDipolar-Blakie-Ferlaino}.
Our goal is to design $a_{s}(t)$, such that the system is driven from an initial ground state with $a_{s}(0)=\asi$ to a final ground state at $a_{s}(t_{f})=\asf$, and within a given control time $t_{f}$.
We first design control schemes within the superfluid regime, i.e. $\asi > \asf > a_{s}^c$, and then design control schemes that cross the superfluid to supersolid transition, $\asf < a_{s}^c < \asi$.

The highly non-linear nature of dipolar BECs makes time-dependent control challenging \cite{2015-OptimalControlBose-Mennemann-Langen,2019-OptimalControlSelfbound-Mennemann-Mauser}.
Practically useful control schemes should minimize excitations after the control time $t_f$, and be simple to engineer in an experimental context.
To measure the performance of control schemes, we define the time-dependent target-overlap fidelity as $F_T(t)=\fabsq{\braket{\Psi_{T}}{\Psi(t)}}$, where $\ket{\Psi_T}$ is the target state, i.e. the ground state of the system with $a_s=\asf$ and $\ket{\Psi(t)} = U(t,0)\ket{\Psi_0}$ is the time-evolved state of the system from the initial ground state $\ket{\Psi_0}$ with $a_s(0)=\asi$ ($U$ denotes the time evolution operator).

In the supersolid regime, there can exist states with energy very close to the ground state, but with an antipodal configuration of peaks within the condensate.
This would lead to $F_T$ being a poor measure of performance, however in the parameter range considered here, the configuration of the peaks is conserved for the control schemes considered.
An extension to this work would be to investigate control to almost degenerate low-energy states, which would be more relevant in quasi-2D trapped dipolar BECs \cite{2021-PatternFormationQuantum-Hertkorn-Pfau,2024-SupersolidformationtimeShortcutExcitation-Alana-Alana,2026-QuantumStabilizedStatesMagnetic-Chomaz}.

Given that the eGPE is a mean-field theory and an approximation to the real-world dynamics encountered in experiments, we focus on control schemes that maximize $F_{T}(t\ge t_{f})$, while minimizing any time-dependent excitations that occur after $t_{f}$.
In order to provide a simple but strict measure of overall control performance for a given $t_f$, we introduce
\begin{align}\label{eq:F_min}
F_{{\min}} = \min\limits_{t_{f} \le t \le 2 t_{f}} F_T(t),
\end{align}
where $2 t_f$ is chosen as a practical timescale for use in experiments.
As will be shown later, changing $a_s(t)$ in time results in density currents within the condensate, which create oscillations in the resulting time-dependent fidelity $F_T(t)$.
Since $F_{\text{min}}$ highlights the worst performance due to these oscillations, it provides a strict measure of control performance, ensuring that the derived control schemes are of high practical value.

\section{Ground state transfer in the superfluid phase}
\label{sec:GSSFSF}

\subsection{\label{sec:var_sf_to_sf} VAIE applied in the superfluid regime}

We first consider ground state transfer in the superfluid regime, from an initial scattering length $a_{s}(t)=\asi$ to a final value $a_{s}(t_{f})=\asf$, in a range larger than the superfluid-supersolid transition ($\asi > \asf > a_{s}^c$).
For a review of the variational approach applied to quantum systems in a number of contexts see \cite{2019-TheoryVariationalQuantum-Yuan-Benjamin}, and for a review of VAIE see \cite{2013-Chapter2Shortcuts-Torrontegui-Mugaa}.

We restrict the system to a cylindrical trap geometry with $\omega_{z}=\omega_{y}\equiv\omega_{\rho}$ and assume a Gaussian ansatz, separable in the axial $x$ and transverse $\rho = (y,z)$ coordinates, given by
\begin{eqnarray}\label{eq:ansatz_cyl}
\Psi_A(x,\rho,t) &=&
C_{0}(t)\exp\left[-\frac{x^{2}}{2 w_{x}^{2}(t)} + ix^{2}\alpha_{x}(t) \right] \times \\
&&\exp\left[-\frac{\rho^{2}}{2 w_{\rho}^{2}(t)} + i\rho^{2}\alpha_{\rho}(t) \right], \notag
\end{eqnarray}
where $C_0(t) = 1/\left( \pi^{3/2} w_{x} w_{\rho}^{2}\right)^{1/2}$.
The time-dependent parameters of the ansatz are $\{\alpha_{x}, \alpha_{\rho}, w_{x}, w_{\rho}\},$ and we drop this explicit time dependence for notational clarity.

\begin{figure*}
\subfloat{%
\includegraphics[width=.32\linewidth]{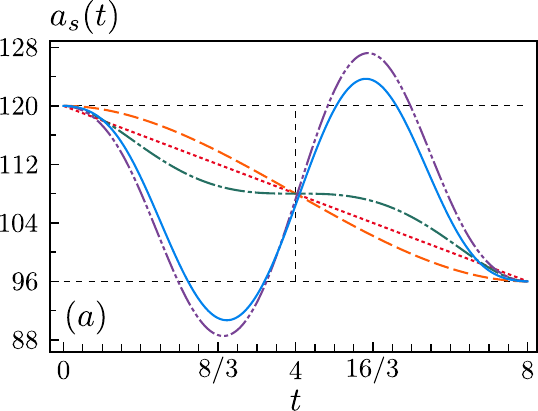}%
}\hspace{.1cm}
\subfloat{%
\includegraphics[width=.32\linewidth]{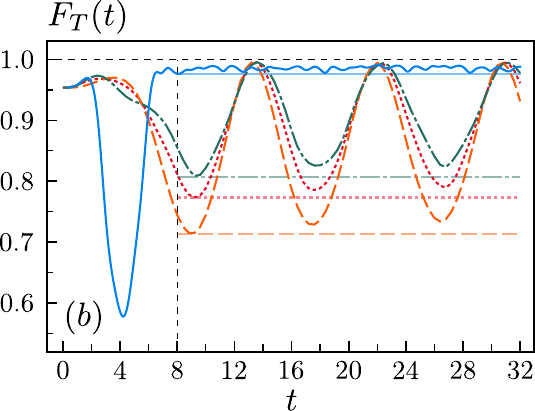}%
}\hspace{.1cm}
\subfloat{%
\includegraphics[width=.32\linewidth]{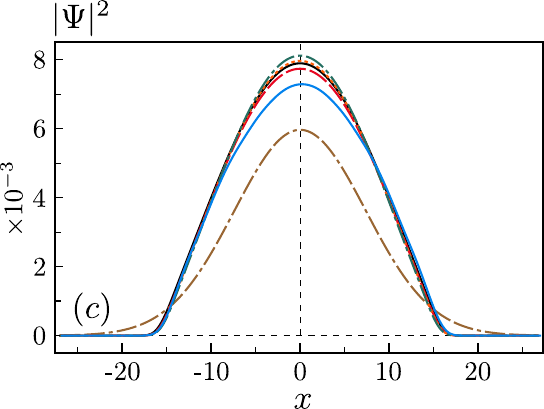}%
}
\caption{Control within the superfluid regime: (a) Examples of $a_s(t)$ for $t_f=8$: three polynomial control schemes defined in \Eqref{eq:traj_defns}, $a_{s}^{(1)}(t)$ (dotted-red), $a_{s}^{(2)}(t)$ (dashed-orange), and $a_{s}^{(3)}(t)$ (dot-dashed green), the VAIE scheme $a_{s}^{(v)}(t)$ (solid blue) and the numerical optimized VAIE scheme $a_{s}^{(vn)}(t)$ (dash-double-dot purple). (b) Plot of $F_{T}(t)$ for $t_f=8$ for the control schemes shown in (a). The vertical dashed-line at $t=8$ denotes the control time $t_f$, with $a_{s}(t)=\asf$ for $t \ge t_{f}$. The horizontal lines for $t \ge t_{f}=8$ highlight the calculation of $F_{{\min}}$ {[}\Eqref{eq:F_min}{]} for each control scheme. (c) Comparison of $x$-axis density slice of $\fabs{\Psi}^2$ at $t_f=8$ for each control scheme, and the exact ground state (solid black) with $\asf=96 \, a_0$. Also shown is the $x$-axis density slice of the ansatz $\Psi_A$ (dash-double-dot brown) \Eqref{eq:ansatz_cyl}, with parameters found by minimizing the energy functional associated with \Eqref{eq:L_ss_ss}.}\label{fig:plot-2}
\end{figure*}

Inserting the ansatz into the Lagrangian density \Eqref{eq:lag_dim}, and integrating over space results in an analytic Lagrangian \cite{2001-TrappedCondensatesAtoms-Yi-You,2016-SelfboundDipolarDroplet-Baillie-Blakie,2018-VorticesSelfboundDipolar-Cidrim-Macri} given by
\begin{align}\label{eq:L_ss_ss}
L &=
\frac{1}{4 w_{x}^{2}}+w_{x}^{2} \left(\alpha_{x}^{2}+\frac{\dot{\alpha}_{x}}{2}+\frac{\omega _{x}^{2}}{4}\right)
\\
&+
\frac{1}{2 w_{\rho}^{2}}+w_{\rho}^{2} \left(2 \alpha_{\rho}^{2}+\dot{\alpha}_{\rho}+\frac{\omega_{\rho}^{2}}{2}\right)
+
\frac{a_{dd} }{2} F\left( \kappa \right) W
\nonumber \\
&+
a_s \: W
+
c_0 \: W^{3/2} \sqrt{a_s} \left( 2 a_s^{2} + 3 a_{dd}^2 \right),\nonumber
\end{align}
where $W = N/ \left( \sqrt{2 \pi} \: w_{x} w_{\rho}^{2} \right)$, $c_{0} = 512 / 75\pi \times  \sqrt[4]{2} /\sqrt{5}$ and $\kappa = w_{\rho} / w_{x}$ with (see Appendix \ref{app:var_dd_term} for details)
\begin{align}\label{eq:F_kappa}
F\left( \kappa \right) =
\frac{1 + 2 \kappa^{2}}{1 - \kappa^{2}} -
\frac{3 \kappa^{2} \arctan \left( \sqrt{1 - \kappa^{2}} \right)}{\left( 1 - \kappa^{2} \right)^{3/2}}.
\end{align}
For each time dependent parameter $q_j \in \{\alpha_{x},\alpha_{\rho},w_{x},w_{\rho}\}$, there is an associated Euler-Lagrange equation,
\begin{align}\label{eq:var-euler-lagrange}
\frac{\partial L}{\partial q_j} - \frac{d}{dt} \left( \frac{\partial L}{\partial \dot{q}_j} \right) = 0,
\end{align}
and for $\alpha_{x}$ and $\alpha_{\rho}$ one obtains (see Appendix \ref{app:euler-lagrange-details})
\begin{align}\label{eq:var-alpha-x-rho}
\alpha_{x} = \frac{\dot{w}_{\rho}}{2 w_{\rho}}, \:
\alpha_{\rho} = \frac{\dot{w}_{x}}{2 w_{x}}, 
\end{align}
Using \Eqref{eq:var-alpha-x-rho} and it's time-derivative, the Euler-Lagrange equations for $w_{x}$ and $w_{\rho}$ can be simplified to
\begin{align}
\ddot{w}_{x} + \omega_{x}^{2} w_{x} &= \frac{1}{w_{x}^3} 
+ \frac{W}{w_{x}}
\Big[ 2 a_s - a_{dd} \: h \left( \kappa \right) +
\nonumber\\
& 3 c_0 \sqrt{W} \sqrt{a_s} \left( 2 a_s^{2} + 3 a_{dd}^2 \right)
\Big],
\label{eq:wx_var} \\
\ddot{w}_{\rho} + \omega_\rho^{2} w_{x} &= \frac{1}{w_{\rho}^3} 
+ \frac{W}{w_{\rho}}
\Big[ 2 a_s - a_{dd} \: g \left( \kappa \right) +
\nonumber \\
& 3 c_0 \sqrt{W} \sqrt{a_s} \left( 2 a_s^{2} + 3 a_{dd}^2 \right)
\Big],\label{eq:wro_var}
\end{align}
where
\begin{align}
g(\kappa)&=\frac{2-7\kappa^{2} + \left[-4 + 9 \, d(\kappa)\right]\kappa^{4} }{(1-\kappa^{2})^{2}},
\nonumber \\
h(\kappa)&=\frac{1+\left[10 - 9\,d(\kappa) \right]\kappa^{2}-2\kappa^{4}}{(1-\kappa^{2})^{2}},
\nonumber \\
d(\kappa)&=\frac{\arctanh{\sqrt{1-\kappa^{2}}}}{\sqrt{1-\kappa^{2}}}.
\end{align}
Scaling \Eqref{eq:wx_var} by $w_{x}$ and \Eqref{eq:wro_var} by $w_{\rho}$, the sum of these equations eliminates $a_{s}(t)$, giving a consistency equation in terms of $w_{x},w_{\rho}$ alone,
\begin{align}\label{eq:consis_equ}
w_{x} \ddot{w}_{x} - w_{\rho}\ddot{w}_{\rho} =
&\frac{1}{w_{x}^{2}} - \frac{1}{w_{\rho}^{2}} +
\omega_{\rho}^{2} w_{\rho}^{2} - w_{x}^{2} \omega_{x}^{2}
\\
&-a_{dd} W \left[ h\left(\kappa\right) - g\left(\kappa\right) \right],
\nonumber
\end{align}
and taking the difference of these scaled equations gives a second equation in terms of $a_s,w_{x},w_{\rho}$,
\begin{align}\label{eq:as_equ}
w_{x} \ddot{w}_{x} + &{w}_{\rho} \ddot{w}_{\rho} =
\frac{1}{w_{x}^{2}} + \frac{1}{w_{\rho}^{2}} -
\omega_{\rho}^{2} w_{\rho}^{2} - w_{x}^{2} \omega_{x}^{2}
\\
&\hspace{1.35cm}-a_{dd} W \left[ h\left(\kappa\right) + g\left(\kappa\right) \right]
\nonumber \\
&+ 2 W \left[
2 a_s +
3 c_0 \sqrt{W} \sqrt{a_s} \left( 
2 a_s^2 + 3 a_{dd}^2
\right)
\right].
\nonumber
\end{align}
To generate practical smooth-control schemes, we enforce $\dot{a}_{s}(t)=0$ for $t=0,t_{f}$, and using \Eqref{eq:consis_equ} and \Eqref{eq:as_equ}, $w_{x}$ and $w_{\rho}$ must satisfy the following boundary conditions
\begin{eqnarray}\label{eq:as_bound_conds}
w_{x}(0)  &&= \wxi,
w_{x}(t_{f}) = \wxf,
w_{\rho}(0)  = \wri,
w_{\rho}(t_{f}) = \wrf,
\nonumber \\
\dot{w}_{x}(0) &&= \dot{w}_{\rho}(0) = \dot{w}_{x}(t_{f}) = \dot{w}_{\rho}(t_{f}) = 0, \nonumber \\
\dddot{w}_{x}(0) &&= \dddot{w}_{\rho}(0) = \dddot{w}_{x}(t_{f}) = \dddot{w}_{\rho}(t_{f}) = 0.
\end{eqnarray}
The values of $ \wxi, \wri$ and $\wxf, \wrf$ are obtained by numerically minimizing the time-independent energy functional associated with \Eqref{eq:L_ss_ss}, for $\asi$ and $\asf$ respectively.

At this point, the Euler-Lagrange equations \Eqref{eq:var-euler-lagrange} have been reduced to \Eqref{eq:consis_equ} and \Eqref{eq:as_equ}, and the boundary conditions.
In general, convenient forms for $w_x(t)$ and $w_\rho(t)$ that satisfy \Eqref{eq:consis_equ} and the boundary conditions can be chosen, and then \Eqref{eq:as_equ} can be solved for $a_s(t)$.
Note that \Eqref{eq:as_equ} is a polynomial in $\sqrt{a_{s}}$ where the coefficients are in terms of the chosen $w_{\rho},w_{x}$ and known constants of the system (see Appendix \ref{app:euler-lagrange-details}).

We choose simple polynomial ans\"atze for $w_{\rho}$ and $w_{x}$, with two free parameters in each polynomial.
While this choice does not satisfy \Eqref{eq:consis_equ} directly, a numerical approximation to the time-integral over $[0,t_f]$ of \Eqref{eq:consis_equ} is minimized for the four free parameters of $w_{\rho}(t)$ and $w_{x}(t)$, such that $w_{\rho}(t)$ and $w_{x}(t)$ are within machine precision of solving \Eqref{eq:consis_equ}.

An alternative more numerically intensive approach would be to assume an analytic form for $w_{\rho}$ or $w_{x}$ that satisfies \Eqref{eq:as_bound_conds}, and then numerically solve for the other width function using \Eqref{eq:consis_equ}. 
Using these $w_{\rho}$ and $w_{x}$, \Eqref{eq:as_equ} can be numerically solved for $a_{s}(t)$.
We mention this to highlight that there are a number of ways to proceed once the simplified Euler-Lagrange equations have been obtained.

\subsection{Results}

For numerical analysis we consider $^{162}$Dy atoms with magnetic moment $\mu = 9.93 \mu_{B}$ ($\mu_{B}$ is the Bohr magneton), consistent with several experiments that have demonstrated time-dependent control of this species \cite{2022-DipolarPhysicsReview-Chomaz-Pfau,2019-ObservationDipolarQuantum-Tanzi-Modugno}. 
We use similar parameters from these experiments, where $N=3\times 10^{4}$ is the number of atoms, and the system is in a cylindrical trapping potential $V(\ve{r})$, with $(\omega_{x},\omega_y,\omega_z)=2\pi\times(15,100,100)$ Hz.
Our goal is to drive the system from a ground state and initial scattering length $a_{s}(0)=120 \,a_{0}$ to a ground state with $a_{s}(t_{f})=96 \,a_{0},$ where $a_{0}$ is the Bohr radius, and $t_f$ is the control time.
Throughout the paper we refer to values of $a_s$ in units of $a_{0}$.

In this subsection, we consider and compare 3 approaches; polynomial schemes given by $a_s^{(1)}(t), a_s^{(2)}(t), a_s^{(3)}(t)$, the VAIE protocol $a_s^{(v)}$ and a numerically optimized VAIE protocol $a_s^{(vn)}$.
The polynomial schemes are given by,
\begin{align}\label{eq:traj_defns}
a_s^{(1)} \left( t \right) &=
\asi  -\tau \, \delta a_s  
\nonumber \\
a_s^{(2)} \left( t \right) &=
\asi + 
 \tau^{2} \left( 2 \tau - 3\right)\delta a_s 
\nonumber \\
a_s^{(3)} \left( t \right) &=
\asi + \tau^{2} 
\left( 24 \tau^3 - 60 \tau^{2} + 50 \tau - 15 \right)  \delta a_s,
\end{align}
where $\tau = t / t_{f}$ and $\delta a_s =  \asi - \asf$.
These control schemes are inspired by adiabatic schemes, and in the limit of large $t_f$, they approach adiabaticity (see Appendix \ref{app:adiab-numerics}) \cite{2019-ShortcutsAdiabaticityConcepts-Guery-Odelin-Muga}.
The $a_s^{(1)}(t)$ scheme is a standard ramp with discontinuous first derivatives at the boundary times $t=0,t_{f}$, $a_s^{(2)}(t)$ is a polynomial with zero first derivatives at the boundaries, and $a_s^{(3)}(t)$ has zero first derivatives at the boundaries and a zero first derivative at $t=t_{f}/2.$
\begin{figure}
\centering
\includegraphics[width=0.95\linewidth]{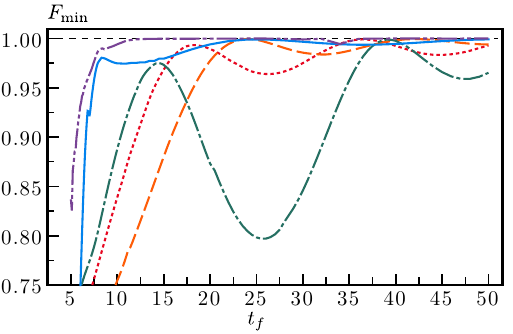}
\caption{Comparison of $a_s(t)$ control schemes in the superfluid regime: $F_{{\min}}$ {[}\Eqref{eq:F_min}{]} versus different control times $t_f$ is calculated for the polynomial schemes, $a_s^{(1)}(t)$ (dotted-red), $a_s^{(2)}(t)$ (dashed-orange), $a_s^{(3)}(t)$ (dot-dashed green), the VAIE scheme $a_s^{(v)}(t)$ (solid blue) and the numerically optimized VAIE scheme (dash-double-dot purple).}\label{fig:plot-3}
\end{figure}

The VAIE protocol $a_s^{(v)}$ was described in the previous section, and we also consider a numerically optimized VAIE scheme $a_s^{(vn)}(t) = a_s^{(v)}(t) + \Delta a_s(t, \lambda_1, \lambda_2)$, which is optimized by a simple search over $\lambda_1$ and $\lambda_2$ ($\Delta a_s$ is defined in Appendix \ref{app:adiab-numerics}).
Examples of the control schemes for $t_f=8$ are shown in Fig. \ref{fig:plot-2} (a).

In Fig. \ref{fig:plot-2} (a) the similarity between the VAIE scheme (solid-blue) and the numerically optimized VAIE scheme (dash-double-dot purple) is apparent, demonstrating that the VAIE scheme is already a near-optimal scheme in this parametrization, as well as an efficient starting point for further optimization.
For example, if $a_s^{(0)}(t)$ (dotted red) was chosen, the optimization procedure as outlined in  Appendix \ref{app:adiab-numerics} would take much longer to converge to very high values of $F_{\text{min}}$.
In practice, for $t_f \gtrsim 8$ the $a_s^{(v)}(t)$ scheme already produces $F_{\text{min}} \gtrsim 0.95$, illustrating that minimal excitation of the system is already achieved in a short time ($\approx 20$ ms).

In Fig. \ref{fig:plot-2} (b), $F_T(t)$ is shown for each of the control schemes, with $0 \le t \le 4 \: t_f$ and $t_f=8$.
For clarity, the numerically optimized scheme $a_s^{(vn)}(t)$ is omitted as it produces very similar results to $a_s^{(v)}(t)$.
Also highlighted is the calculation of $F_{\text{min}}$, denoted by the horizontal lines associated with each control scheme {[} \Eqref{eq:F_min}{]}.
Large excitations are generated within the condensate using the polynomial schemes, resulting in oscillations of $F(t)$ which continue well beyond the control time $t_f=8$.
The $a_s^{(v)}(t)$ scheme performs significantly better, with minimal excitation after $t_f=8$.
Note that the numerically optimized $a_s^{(vn)}(t)$ scheme improves upon the $a_s^{(v)}(t)$ scheme, but in this example the difference is minimal, also demonstrating the effectiveness of the VAIE approach even without optimization.
\begin{figure}
\flushleft 
\includegraphics[width=0.99\linewidth]{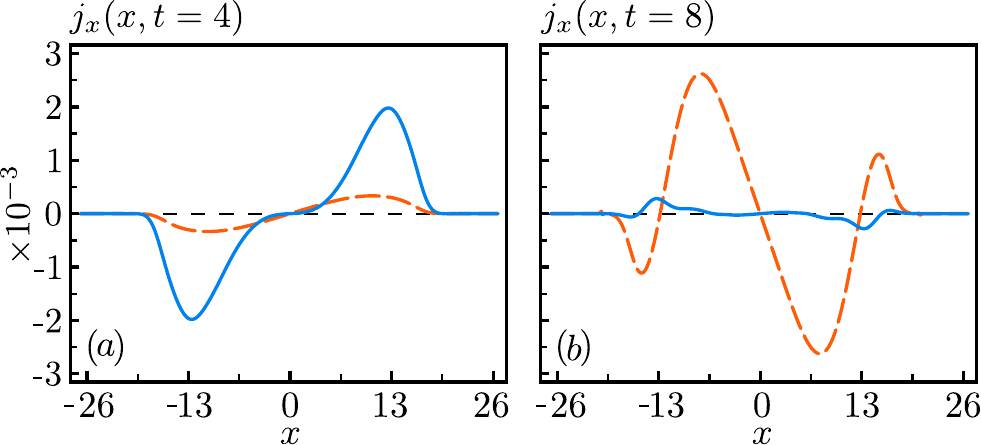}
\caption{ Snapshots of $j_{x}(x,t)$ {[}\Eqref{eq:jx_defn}{]} with $t_{f}=8$, the polynomial scheme $a_s^{(2)}(t)$ (dashed-orange) and the VAIE scheme $a_s^{(v)}(t)$ (solid blue). (a) $t=4$ and (b) $t=t_f$ demonstrate that $a_s^{(v)}(t)$ produces an initially larger current density flow than $a_s^{(2)}(t)$, and this flow is suppressed as $t \rightarrow t_{f}$ and beyond, as shown in (b) at $t=t_f$.
This is in contrast to $a_s^{(2)}(t)$, which initially has lower current density flow, but results in large oscillations in the current density for $t \ge t_f$.}
\label{fig:plot-4}
\end{figure}

In Fig. \ref{fig:plot-2} (c) the $x$-axis density slice at $t_f=8$ is shown for the different control schemes, the numerical ground state and the variational ansatz.
While the different schemes agree reasonably well with the exact ground state at $t=t_f$, the practical performance of the schemes is highlighted more appropriately by using $F_{\text{min}}$, as highlighted in Fig. \ref{fig:plot-3}, where $F_{\text{min}}$ is compared for each of the control schemes.

The VAIE scheme $a_s^{(v)}(t)$ (solid blue) offers a significant improvement over the three polynomial schemes $a_s^{(1)}(t)$ (dotted-red), $a_s^{(2)}(t)$ (dashed-orange), $a_s^{(3)}(t)$ (dot-dashed green).
The numerically optimized VAIE scheme $a_s^{(vn)}(t)$ (dash-double-dot purple) offers some further improvement, but at the cost of additional computation.

As previously discussed, the system is highly susceptible to excitations induced by a  time-dependent $a_{s}(t)$, since it directly affects the contact and LHY terms in \Eqref{eq:lag_dim} \cite{2011-QuantumFluctuationsDipolar-Lima-Pelster}. This issue is compounded by the non-local dipolar interaction, and the net effect of these interactions in this trapping geometry is to create an anisotropic space and time-dependent phase.
As a result, currents develop in the condensate density, even for approximately adiabatic control schemes (see Appendix \ref{app:adiab-numerics}).
\begin{figure}
\vspace{0.3cm}
\includegraphics[width=0.99\linewidth]{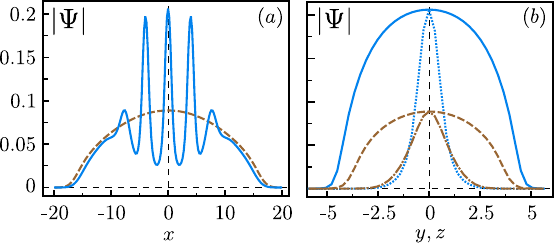}\\
\includegraphics[width=0.99\linewidth]{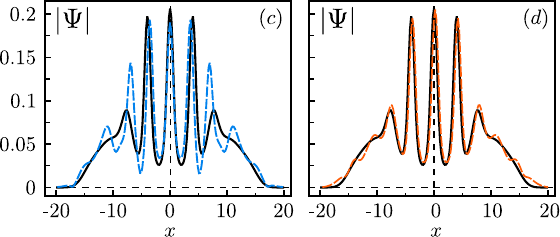} 
\caption{(a)-(b) Axis density slices of the exact groundstates at initial and final $a_s$: (a) Numerical ground state $x$-axis density slices $|\Psi(x,0,0)|$ for $\asi = 95.9 \, a_0$ (dashed-brown) and $\asf = 90.9 \, a_0$ (solid blue). Note that Fig. \ref{fig:system_setup.jpg} is a 3D representation of the exact ground state at $a_s=90.9 a_0$. (b) Numerical ground state $y$ and $z$-axis density slices for $\asi = 95.9 \, a_0$, $|\Psi(0,y,0)|$ (dot-dashed brown) and $|\Psi(0,0,z)|$ (dashed brown), and for $\asf = 90.9 \, a_0$, $|\Psi(0,y,0)|$ (dotted-blue) and $|\Psi(0,0,z)|$ (solid blue).(c)-(d) Comparison of final-state $x$-axis density slice at $t=t_f=45$ with the exact ground state at $a_s^f = 90.9 \; a_0$ (solid-black): (c) Time-evolved state using polynomial scheme $a_s^{(3)}$ (dashed-blue), and (d) using numerically optimized scheme $a_s^{(n)}(t)$ (dashed-orange). }\label{fig:ss-sf-gs}
\end{figure}

To illustrate this effect, we numerically calculate the time-dependent probability density current $j_{x}(x, t)$ given by 
\begin{align}\label{eq:jx_defn}
j_{x}(x,t) = \frac{\hbar}{m} {\rm Im} \left(\Psi^{*}(x,0,t)\frac{\partial}{\partial x}\Psi(x,0,t) \right).
\end{align}
The improvement in $F_{\min}$ that the VAIE scheme produces can be illustrated by the suppression of current density flow in the $x-$axis.
In Fig. \ref{fig:plot-4}, snapshots of $j_{x}(x,t)$ are shown at $t = 4$ and $t=t_f=8$, obtained by using the polynomial scheme $a_s^{(2)}(t)$ and the VAIE scheme $a_s^{(v)}(t).$ Although initially $a_s^{(v)}(t)$ (solid blue) induces a larger $x$-axis current density flow than $a_s^{(2)}(t)$ (dashed-orange), at the end of the control time $j_{x}(x,t_{f})$ has been suppressed. This is in contrast to the $y,z$ directions (not shown), which experience higher frequency current density flows, and whose effects can be seen as the small amplitude oscillations in $F(t)$ in Fig. \ref{fig:plot-2} (b) (solid blue).

\begin{figure*}[t]
\subfloat{%
\includegraphics[width=.32\linewidth]{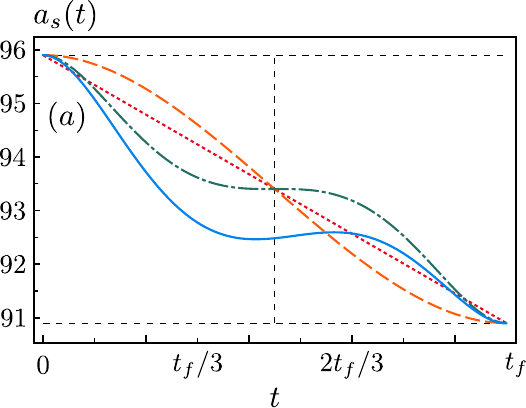}%
}\hspace{.1cm}
\subfloat{%
\includegraphics[width=.32\linewidth]{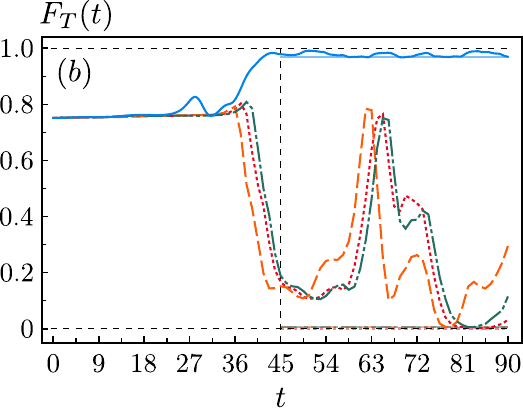}%
}\hspace{.1cm}
\subfloat{%
\includegraphics[width=.32\linewidth]{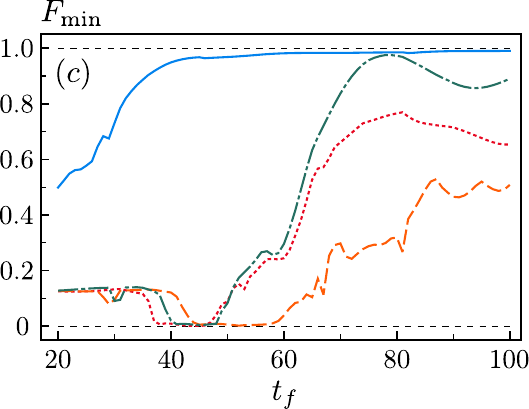}%
}
\caption{Control across the superfluid-supersolid transition: (a) Examples of $a_{s}(t)$ with $t_{f}=45$ and $a_s^{(1)}(t)$ (dotted-red), $a_s^{(2)}(t)$ (dashed-orange), $a_s^{(3)}(t)$ (dot-dashed green), and the numerical optimized scheme $a_s^{(n)}(t)$ (the result of optimizing $a_s^{(A)}(t)$ and $a_s^{(B)}(t)$, both of which converge to $a_s^{(n)}(t)$). (b) Time-dependent fidelity $F_T(t)$ for $t_{f}=45$, with same labeling as (a).
The vertical dashed line at $t_{f}=45$ denotes the end of the control schemes, with $a_{s}(t)=\asf$ for each control scheme after $t \ge t_{f}$.
The horizontal lines for $t \ge t_{f}=45$ highlight the calculation of $F_{{\min}}$ for each control scheme. (c) Plot of $F_{{\min}}$ as defined in \Eqref{eq:F_min} with $t_{f}=45$ and, $a_{s}(t)=a_s^{(1)}(t)$ (dotted-red), $a_s^{(2)}(t)$ (dashed-orange), $a_s^{(3)}(t)$ (dot-dashed green), $a_s^{(n)} (t)$ (solid blue).}\label{fig:sf-ss-fid}
\end{figure*}

\section{\label{sec:var_sf_to_ss} Superfluid to supersolid transition: direct optimization}

As outlined in Sec. \ref{sec:intro}, the ansatz of \Eqref{eq:ansatz_cyl} fails to capture the dynamics across the superfluid-supersolid transition adequately, and instead we adopt a simple optimization scheme based on the polynomial schemes of the last section \cite{2025-KibbleZurekScalingSuperfluidsupersolid-Kirkby-Chomaza,2021-BraggScatteringUltracold-Petter-Ferlaino, 2026-QuantumStabilizedStatesMagnetic-Chomaz,2021-NumericalCalculationDipolarquantumdroplet-Lee-Blakiea}.

Our goal is to drive the system from a ground state with $a_s(0) = 95.9 \,a_0$, across the superfluid-supersolid transition to a ground state at $a_s(t_{f}) = 90.9 \,a_0$.
As shown in Fig \ref{fig:ss-sf-gs} (a), the ground state for $a_s = 90.9 \,a_0$ has a modulated density in the $x$ direction, in contrast to the ground state for $a_s = 95.9 \, a_0$.
In addition to achieving this state transfer, we focus on simple smooth control schemes that could prove useful in experimental applications.

To this end, we define
\begin{align}\label{eq:ss-traj-defns}
a_s^{(A)}(t) = a_s^{(2)}(t) + \Delta a_s (t, \lambda_1, \lambda_2),
\nonumber \\
a_s^{(B)}(t) = a_s^{(3)}(t) + \Delta a_s (t, \lambda_1, \lambda_2),
\end{align}
where $a_s^{(2)}(t)$ and $a_s^{(3)}(t)$ are from Eq. \ref{eq:traj_defns}, and $\Delta a_s$ is defined in Appendix \ref{app:adiab-numerics}.
We numerically optimize $F_{\text{min}}$ with respect to $\lambda_1$ and $\lambda_2$, see Appendix \ref{app:adiab-numerics} for details.
Interestingly, the optimization of $a_s^{(A)} (t)$ and $a_s^{(B)} (t)$ converges to the same trajectory, which we label $a_s^{(n)}(t)$ for clarity.
This suggests that the optimization performed over this two-dimensional parametrization of $\Delta a_s (t, \lambda_1, \lambda_2)$, is near optimal for this problem.

In Fig. \ref{fig:sf-ss-fid} (a), examples of the three polynomial schemes are shown for $t_{f}=45$, $a_s^{(1)}(t)$ (dotted-red), $a_s^{(2)}(t)$ (dashed-orange), $a_s^{(3)}(t)$ (dot-dashed green), as well as the numerically optimized scheme $a_s^{(n)} (t)$ (solid blue).
The resulting time-dependent fidelity is presented in Fig. \ref{fig:sf-ss-fid} (b), with $0 \le t \le 2 t_f$, and $F_{\text{min}} \approx 0.97$ for $a_s^{(n)} (t)$.
The optimized scheme $a_s^{(n)}(t)$ has significant improvement over the polynomial schemes, and in Fig. \ref{fig:sf-ss-fid} (c) this improvement is further demonstrated by the considerable improvement in $F_{\text{min}}$ across a wide range of $t_f$.

Since the transition considered here is continuous \cite{2026-QuantumStabilizedStatesMagnetic-Chomaz,2020-SupersolidityElongatedDipolar-Blakie-Ferlaino, 2025-LocalizedStatesDipolar-Steinberg-Thiele}, a smooth $F_{\text{min}}$ fidelity landscape around $\lambda_1=\lambda_2=0$ is found, although for small $t_f$ the landscape can be complicated and a global maximum for $F_{\text{min}}$ becomes harder to identify.
However, from $t_f\approx 36$, $F_{\text{min}} \ge 0.9$, which is well within the practical timescales of current experimental setups ($\approx 92$ ms) \cite{2019-ObservationDipolarQuantum-Tanzi-Modugno,2019-LongLivedTransientSupersolid-Chomaz-Ferlaino,2019-TransientSupersolidProperties-Bottcher-Pfau}.

\begin{figure}
\centering
\includegraphics[width=0.99\linewidth]{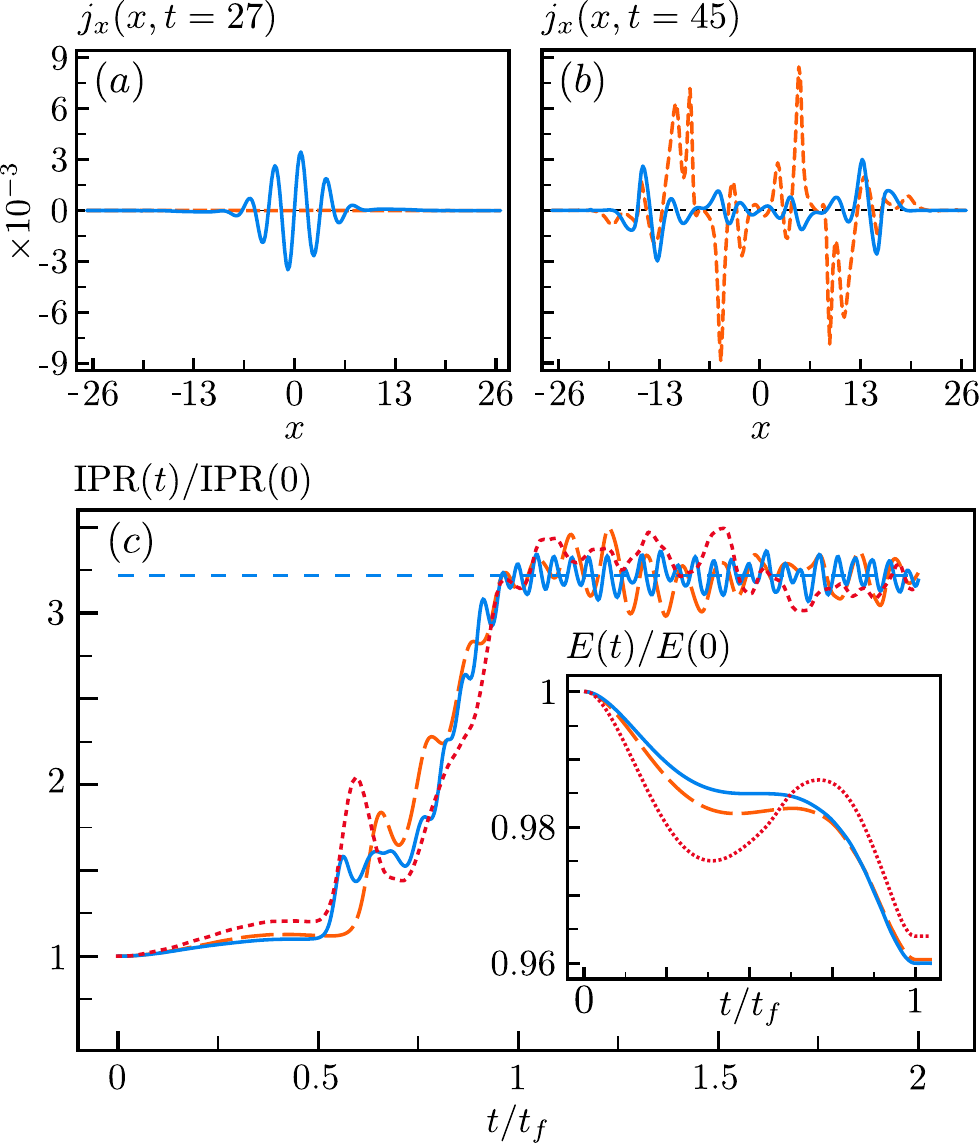}
\caption{Snapshots in time of $j_{x}(x,t)$ {[}\Eqref{eq:jx_defn}{]} for $t_f=45$: (a) Snapshot at $t=27$ and at (b) $t=t_f$, for $a_s^{(2)}(t)$ (dashed-orange) and the numerical scheme $a_s^{(n)}(t)$ (solid blue), showing that the variational scheme has an initially larger current density flow, that is suppressed as $t \rightarrow t_{f}$ and beyond.
This is in contrast to the polynomial scheme $a_s^{(2)}(t)$, which results in continued larger current density oscillatory-flow. (c) IPR for $a_s^{(n)}(t)$ and $t_f=30$ (dotted-red), $t_f=45$ (dashed-orange) and $t_f=90$ (solid-blue). Inset is the time-dependent energy for the same $t_f$.}\label{fig:sf-ss-jx}
\end{figure}

In Fig. \ref{fig:sf-ss-jx} the $x$-axis density at $t=t_f=45$ of the time-evolved state using (a) $a_s^{(2)}(t)$ and (b) $a_s^{(n)}(t)$ is compared to the exact ground state at $a_s^f=90.9 \; a_0$.
While $a_s^{(2)}(t)$ produces large oscillations of the modulated density, these are suppressed by $a_s^{(n)}(t)$.
The improvement of the optimized $a_s^{(n)}(t)$ scheme over the polynomial schemes, can also be seen as a suppression of density currents within the condensate, as illustrated previously in the superfluid regime of Sec. \ref{sec:var_sf_to_sf}.
In Fig. \ref{fig:sf-ss-jx}, an example of this suppression is shown for $t_f=45$.
Fig. \ref{fig:sf-ss-jx} (a) shows an intermediary time of $t=27$, while Fig. \ref{fig:sf-ss-jx} (b) is at $t=t_f=45$.
Initially $a_s^{(n)}(t)$ produces a larger density flow, this is suppressed for $t \ge t_f$.
This larger initial flow can be understood from Fig. \ref{fig:sf-ss-fid} (a), where the optimized scheme (solid blue) drops through the phase transition earlier than the polynomial schemes.
From approximately $t_f/2$, the optimized schemes then raise $a_s$ to a local maximum near $2 t_f / 3$ before smoothly falling to $a_s^f$.
This behavior is somewhat similar to the "parachute-jump method" reported in \cite{2024-SupersolidformationtimeShortcutExcitation-Alana-Alana}, and is typical of fast dynamical control schemes that suppress final excitations, for example STA schemes that implement BEC cooling \cite{2019-ShortcutsAdiabaticityConcepts-Guery-Odelin-Muga}.

To quantify the modulation in the wavefunction density, we use the inverse participation ratio (IPR) defined as \cite{2022-CrossingSuperfluid-supersolid-alana-modugno,2024-SupersolidformationtimeShortcutExcitation-Alana-Alana}
\begin{align}
\text{IPR}(t) = \int d^3r \fabs{\Psi(\ve{r},t)}^4.
\end{align}
In Fig. \ref{fig:sf-ss-jx} (c), the IPR for $a_s^{(n)}(t)$ is shown for $t_f=30$ (dotted-red), $t_f=45$ (dashed-orange) and $t_f=90$ (solid-blue), with $0 \le t \le 2 t_f$, and the time axis scaled by $1/t_f$.
The average of the IPR for $t_f = 90$ from $t_f \le t \le 2 t_f$ is approximately $3.2$ (dashed-blue horizontal line), representing the value of a supersolid state with minimal excitations \cite{2024-SupersolidformationtimeShortcutExcitation-Alana-Alana}.
The small oscillations of the IPR in Fig. \ref{fig:sf-ss-jx} (c) for $t_f=45$ (dashed-orange) are consistent with $F_T$ using $a_s^{(n)}(t)$ in Fig. \ref{fig:sf-ss-jx} (c) (solid-blue).
In Fig. \ref{fig:sf-ss-jx} (c) for $t_f = 30$ (dotted-red), the IPR can be seen to increase after $t_f$ and oscillate about the value 3.2, while $F_{\text{min}} \approx 0.73$ in Fig. \ref{fig:sf-ss-fid} (c) for the same control scheme (solid-blue).

The time-dependent energy {[}energy functional of \Eqref{eq:lag_dim}{]} for $a_s^{(n)}(t)$ is shown in the inset of Fig. \ref{fig:sf-ss-jx} (c), with the same labeling as the larger plot.
As expected, shorter control times require a larger variation in energy, and the energy of the final state is higher as $t_f $ is made shorter \cite{2017-TradeOffSpeedCost-Campbell-Deffner,2017-QuantumSpeedLimits-Deffner-Campbell-S.Campbell}.

\section{Conclusions}
We have developed control schemes for dipolar Bose-Einstein condensates, modeled by the extended Gross-Pitaevskii equation with non-local dipolar interaction term.
Smooth control schemes for the interatomic time-dependent scattering length $a_{s}(t)$ were designed and optimized for a dipolar BEC in a cylindrical harmonic trap, aimed at controlling evolution in the superfluid regime and the superfluid-supersolid transition.

The variational approach with inverse engineering was used in the superfluid regime to design simple and smooth $a_{s}(t)$ that outperformed standard polynomial schemes, without requiring full numerical optimization. 
A direct numerical optimization procedure was used as an extension to this approach, which further improved these results, and also demonstrated that low-dimensionally parameterized optimization schemes can work very well in this system.
These results show that for superfluid to superfluid state transfer, simple and smooth $a_{s}(t)$ can be engineered in a number of ways.

For the superfluid to supersolid transition, the polynomial schemes and a direct numerical optimization were applied.
In this case, a naive application of the variational approach with inverse engineering and a separable Gaussian ansatz is not well suited to the control problem, but a simple two parameter optimization performs well as shown in Fig. \ref{fig:sf-ss-fid} (b) and (c).
The evolution of the superfluid phase and formation of the supersolid correspond with the formation of probability currents, and the improvement due to the optimized schemes leads to their suppression. 

A future avenue of research could be to consider a more complicated ansatz with the VAIE technique, to design control schemes for the phase transition, for example a sum of Gaussians \cite{2010-VariationalMethodsCoupled-Rau-Wunner1,2010-VariationalMethodsCoupled-Rau-Wunner2}.
A difficulty with this approach is solving the resulting Euler-Lagrange equations, but an alternative formulation of the problem using the Dirac-Frenkel-McLachlan variational principle may offer a simpler way to also include inverse engineering \cite{2010-VariationalMethodsCoupled-Rau-Wunner1,2010-VariationalMethodsCoupled-Rau-Wunner2}.
However, the computational effort required to implement such techniques would need to be weighed against the insight gained, considering that for practical purposes the simple method used in this paper can achieve useful control schemes with minimal effort.

The formation time of the supersolid state could also be considered, in the context of quantum speed limits or Kibble-Zurick scaling \cite{2025-KibbleZurekScalingSuperfluidsupersolid-Kirkby-Chomaza, 2017-QuantumSpeedLimits-Deffner-Campbell-S.Campbell}.
Alternative optimization techniques could also be used, for example, quantum control algorithms such as GRAPE or CRAB restricted to a set of smooth or experimentally relevant control-pulse basis functions \cite{2005-OptimalControlCoupled-Khaneja-Reiss,2011-OptimalControlTechnique-Doria-Calarco-ea-S.Montangero, 2019-OptimalControlSelfbound-Mennemann-Mauser}.
However, their computational cost can be much higher, and may need tailoring to specific experimental frameworks.

Another extension of the results presented here would be to investigate the addition of a loss term in the Hamiltonian, in line with known experimental losses during condensate formation and manipulation \cite{2009-PhysicsDipolarBosonic-Lahaye-Pfau, 2020-NewStatesMatter-Bottcher-Pfau,2016-BoseEinsteinCondensation-Pitaevskij-Stringari,2022-DipolarPhysicsReview-Chomaz-Pfau, 2026-QuantumStabilizedStatesMagnetic-Chomaz}.
In addition, a possible future avenue of research could be the investigation of superfluid to supersolid transitions within a time-dependent trap potential, in addition to or separately from the scattering length.

\begin{acknowledgments}
We wish to acknowledge support from IKUR PCI2022-132984, EPIQUS GA 899368 (H2020-FETOPEN-2018-2020), PID2021-126273NB-I00 and PCI2022-132947 projects of MCIN/AEI/10.13039/501100011033 and ERDF “A way of making Europe”, the Basque Government through Grant No. IT1470-22, and by the European Research Council through the Advanced Grant “Supersolids” (No. 101055319).
X.C. also appreciates the Severo Ochoa Centres of Excellence program through Grant EX2024-001445-S and the Spanish Ministry of Economic Affairs and Digital Transformation through the QUANTUM ENIA project call-Quantum Spain project.
\end{acknowledgments}

\appendix

\section{Integral of the dipole interaction term using separable ansatz}\label{app:var_dd_term}
The dipole interaction term in \Eqref{eq:lag_dim} can be calculated analytically when using the ansatz from \Eqref{eq:ansatz_cyl} (for clarity we label this term as $E_{dd} $) \cite{2001-TrappedCondensatesAtoms-Yi-You,2003-ExpansionDipolarCondensate-Yi-You,2006-ExpansionDynamicsDipolar-Giovanazzi-Pfau}, with
\begin{align}
E_{dd} = \frac{N}{2} \int d^3r  \int d^3r' U_{dd} \left( \ve{r}-\ve{r}' \right)\fabsq{\Psi_A(\ve{r}',t)\Psi_A(\ve{r},t)}.
\end{align}
Applying a change of variables $\ve{u} = \ve{r}-\ve{r}'$, and Cartesian coordinates scaled by $w_\rho$, the Gaussian terms can be combined and an integration over $\ve{r}$ results in
\begin{align}
E_{dd} = C_0
\int d^3r \: U_{dd}\left( \ve{r}\right) \: f(\ve{r}),
\end{align}
where $C_0 = N / (4 \sqrt{2 \pi^3} w_\rho^2 w_x)$ and 
\begin{equation}
f(\ve{r}) = \exp \left[-\frac{1}{2}\left(\kappa x^2 + y^2 + z^{2} \right)\right],
\end{equation}
$\kappa = w_\rho / w_x$ and $U_{dd}$ is from \Eqref{eq:U_dip}.
Using the definition of the Dirac delta distribution, and the known Fourier transforms of $U_{dd}(\ve{r})$ \cite{2022-DipolarPhysicsReview-Chomaz-Pfau} and $f(\ve{r})$, the integral can be written
\begin{align}
E_{dd} &= C_0 
\int d^3r \: U_{dd}\left( \ve{r}\right) \: f(\ve{r})
\nonumber \\
&=C_1 \int d^3r
\left[ \cF^{-1} \left( \widetilde{U}_{dd} \right) \right]
\left[ \cF^{-1} \left( \widetilde{f} \right) \right]
\nonumber \\
&=C_1 \int \frac{d^3k}{\left(2 \pi \right)^3} 
\widetilde{U}_{dd} \left( \ve{k} \right) 
\widetilde{f} \left( -\ve{k} \right),
\end{align}
where $C_1 = C_0 / \kappa \sqrt{2 / \pi} $.
The integral in cylindrical coordinates becomes
\begin{align}
E_{dd} &= C_2
\int d^3k
\left( \frac{k_\rho^{2}\sin^{2} k_{\phi} }{k_\rho^{2} + k_{x}^{2}} - \frac{1}{3} \right)
k_\rho 
\exp \left(
- \frac{k_{x}^{2}}{2 \kappa^{2}}
- \frac{k_\rho^{2}}{2}
\right),
\end{align}
with $C_2=C_1 C_{dd}$.
This integral can be done analytically, resulting in
\begin{align}\label{app:analytic_dip_int}
E_{dd} &= C_3
\left[
\frac{1+2\kappa^2}{ 1 - \kappa^2} -
\frac{3 \kappa^2 \arctanh{\sqrt{1 - \kappa^2}}}
{\left( 1 - \kappa^2 \right)^{3/2}}
\right],
\end{align}
 $C_3 = C_{dd}N / (24 \sqrt{2} \pi^{3/2} w_\rho^2 w_x) = W a_{dd}/2$, in agreement with \Eqref{eq:L_ss_ss}.
A similar calculation can be done for $\veuh{z}$-oriented dipoles with the trap extended along the $\veuh{z}-$ axis rather than the $\veuh{x}-$ axis, resulting in \eqref{app:analytic_dip_int} scaled by $-2$ \cite{2006-ExpansionDynamicsDipolar-Giovanazzi-Pfau}.

\section{Solving for $a_{s}(t)$ using the Euler-Lagrange equations}\label{app:euler-lagrange-details}
To solve for $a_{s}(t)$, we write the terms from \Eqref{eq:lag_dim} as
\begin{align}\label{eq:app_lag_terms}
\cL_T(\ve{r},t) &=  
\frac{i}{2} \left(
\Psi   \frac{\partial \Psi^{*}}{\partial t}  - 
\Psi^* \frac{\partial \Psi}{\partial t}
\right)
=
\text{Im}\left(
\Psi^* \frac{\partial \Psi}{\partial t}
\right), \\
\cL_0(\ve{r},t) &= \frac{1}{2} \fabsq{\nabla \Psi(\ve{r},t)} +
V(\ve{r}) \fabsq{\Psi(\ve{r},t)} +
\nonumber \\ 
&
\frac{N}{2} \int d^3r' U_{dd} \left( \ve{r}-\ve{r}' \right) \fabsq{\Psi(\ve{r}',t)}
\fabsq{\Psi(\ve{r},t)},
\nonumber \\
\cL_{a}(\ve{r},t) &= \frac{1}{2} g_s\left( t\right) \fabs{\Psi(\ve{r},t)}^4
+ \frac{2}{5}  g_{L}\left( t\right) \fabs{\Psi(\ve{r},t)}^{5}
\nonumber
\end{align}
and define $L_\nu \equiv \int d^3r \: \cL_\nu$, with the Euler-Lagrange equations for the ansatz parameters $q_i$ given by
\begin{align}\label{eq:app_EL}
\frac{\partial L_{a}}{\partial q_i} +
\frac{\partial }{\partial q_i} \left( L_{T} + L_{0}\right)
- \frac{d}{dt} \frac{\partial L_{T}}{\partial \dot{q}_i} = 0,
\end{align}
since $\Psi = \Psi[\ve{r},t,q_j(t)]$ and only $L_T$ is a function of $\dot{q}_i$.
Recalling the definitions of \Eqref{eq:contact_interaction} and \Eqref{eq:LHY_interaction}, and setting 
\begin{align}
&C_0 = \frac{256}{16} \sqrt{\pi} N^{3/2},
&C_1 = 2 \pi N &
&C_2 = \frac{3}{2} a_{dd}^2 \,C_0,
\end{align}
the Lagrangian term $L_{a}$ can be written as:
\begin{align}
L_{a} = C_0 I_{5}a_s^{5/2}(t) + C_1 I_{4}a_s(t) + C_2 I_{5}a_s^{1/2}(t),
\end{align}
with
$I_4 \equiv \int d^3r \fabs{\Psi}^4$ and $I_5 \equiv \int d^3r \fabs{\Psi}^5$.
Hence the Euler-Lagrange equations \Eqref{eq:app_EL} can be written
\begin{align}\label{app:as_poly}
d_0^{(j)}a_s^{5/2}(t) +
d_1^{(j)}a_s(t) +
d_2^{(j)}a_s^{1/2}(t) + d_3^{(j)} = 0,
\end{align}
where,
\begin{align}
d_0^{(j)} &= C_0 \ \frac{\partial I_5}{\partial q_j}, \quad
d_1^{(j)} = C_1 \; \frac{\partial I_4}{\partial q_j}, \quad
d_2^{(j)} = C_2 \; \frac{\partial I_5}{\partial q_j},
\nonumber \\
d_3^{(j)} &= 
\frac{\partial L_0}{\partial q_j} +
\frac{\partial L_T}{\partial q_j} -
\frac{d}{dt} \frac{\partial L_{T}}{\partial \dot{q}_i}.
\end{align}
\Eqref{app:as_poly} is a polynomial in $a_s(t)$ and can be solved numerically.
Note that $d_k^{(j)}(t)$, which makes solving for $a_s(t)$ challenging for more than a few $q_j$.
The following relations can simplify the evaluation of \Eqref{app:as_poly}
\begin{align}
&\frac{\partial I_4}{\partial q_j} = \int d^3r \frac{\partial }{\partial q_j}
\left( \Psi^* \Psi\right)^2 = 
4 \int d^3r \fabsq{\Psi} \text{Re} \left( \Psi^* \frac{\partial \Psi}{\partial q_j} \right),
\nonumber \\
&\frac{\partial I_5}{\partial q_j} = \int d^3r \frac{\partial }{\partial q_j}
\left( \Psi^* \Psi\right)^{5/2} = 
5 \int d^3r \fabs{\Psi}^3 \text{Re} \left( \Psi^* \frac{\partial \Psi}{\partial q_j} \right),
\nonumber \\
&\frac{\partial L_{T}}{\partial q_j} =
\int d^3r \:\text{Im} \left[
\frac{\partial}{\partial q_j} \left( \Psi^* \dot{\Psi} \right)
\right].
\end{align}
An additional useful constraint on the choice of ansatz is to assume a form where the phase is not space dependent, as this simplifies solving the resulting Euler-Lagrange equations \cite{1997-VariationalThomasFermiTheory-Timmermans-Huang}.
However, the transition from superfluid to supersolid involves the development of a space and time-dependent phase, and alternative ans\"atze are more appropriate \cite{2022-DipolarPhysicsReview-Chomaz-Pfau,2026-QuantumStabilizedStatesMagnetic-Chomaz}.

In the case of the superfluid to superfluid control, the separable ansatz introduced in Section \ref{sec:var_sf_to_sf} and Lagrangian \Eqref{eq:L_ss_ss}, result in coupled equations that can be simplified.
Specifically, to obtain \Eqref{eq:var-alpha-x-rho} from \Eqref{eq:var-euler-lagrange}, one starts with
\begin{align}
\frac{\partial L}{\partial \alpha_{x}} 
- \frac{d}{dt} \left( \frac{\partial L}{\partial \dot{\alpha}_{x}}\right)&= w_{x} \left( \alpha_{x} w_{x} - \dot{w}_{x} \right) = 0,
\nonumber \\
\frac{\partial L}{\partial \alpha_{\rho}} 
- \frac{d}{dt} \left( \frac{\partial L}{\partial \dot{\alpha}_{\rho}}\right)&= 2 w_{\rho} \left( \alpha_{\rho} w_{\rho} - \dot{w}_\rho \right) = 0,
\end{align}
which results in the expressions of \Eqref{eq:var-alpha-x-rho}.
Combining \Eqref{eq:var-alpha-x-rho} and using $g_s$ and $g_L$ [\Eqref{eq:contact_interaction} and \Eqref{eq:LHY_interaction}], the equation for $a_{s}(t)$ [\Eqref{eq:as_equ}] becomes a polynomial in $a_{s}(t)$ as shown in \Eqref{app:as_poly} where $d_0,\dots,d_3$ are in terms of $w_{x}(t)$ and $w_{\rho}(t)$.
Then there is freedom to choose $w_{x}(t)$ and $w_{\rho}(t)$, such that \Eqref{eq:consis_equ} and the boundary conditions \Eqref{eq:as_bound_conds} are satisfied, and $a_s(t)$ can be solved for using \Eqref{app:as_poly}.

\section{Adiabatic limit and numerics }\label{app:adiab-numerics}
\begin{figure}
\centering
\includegraphics[width=1\linewidth]{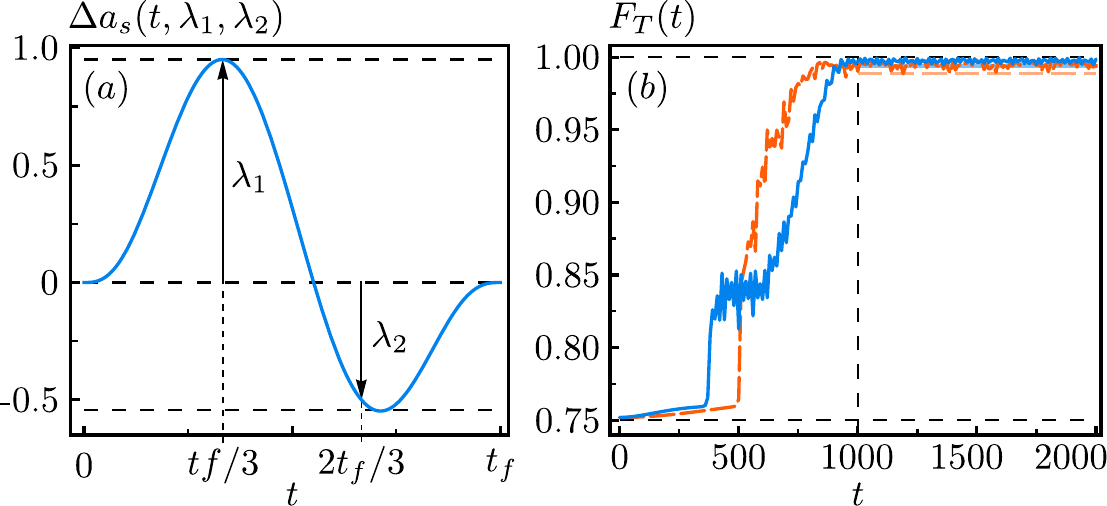}
\caption{(a) Examples of quasi-adiabatic schemes in the large $t_{f}$ limit, $a_s^{(2)}(t)$ (solid-blue) and $a_s^{(3)}(t)$ (dashed-orange). As with the other results in this work, a minimum (dimensionless) time step of $\Delta \tau = 10^{-3}$ is used. (b) Diagram of $\Delta a_s(\tau, \lambda_1, \lambda_2)$, used to optimize the polynomial schemes of \Eqref{eq:ss-traj-defns}.}\label{fig:app}
\end{figure}

To illustrate the general behavior of the system in the limit of large $t_{f}$, we simulate the system for $t_f=10^{3}$ in Fig. \ref{fig:app} for two example polynomial schemes.
While the timescales here are beyond current experimental implementation, the purpose here is to illustrate that there does exist an approximate adiabatic limit.

For numerical simulation, the time-dependent results presented in this paper use a dimensionless time step of $\Delta \tau = 1\times 10^{-3}$.
These results were checked against $\Delta \tau = 1\times 10^{-4}$ and $\Delta \tau = 1\times 10^{-5}$, for several different control schemes.
The spatial grid used the following number of points in each spatial direction,  $\left( N_{x},N_y,N_z \right) = \left(256,64,64\right)$, and several results were checked for consistency against a finer grid of $\left( N_{x},N_y,N_z \right) = \left(512,128,128\right)$.
In Appendix \ref{app:dip-term} the details of the implementation of the dipolar interaction term are discussed.
Several publicly available codes exist for simulating this system \cite{2009-FortranProgramsTimedependent-Muruganandam-Adhikari,2015-FortranProgramsTimedependent-Kumar-Adhikari,2016-CUDAProgramsSolving-Loncar-Adhikari,2016-OpenMPOpenMPMPI-Loncar-Balaz,2017-HybridParallelAlgorithms-Loncar-Balaz}, however we developed custom C/CUDA to simulate the system.

For numerical optimization, we use the polynomial schemes {[}\Eqref{eq:traj_defns}{]}as a starting point and add the following polynomial
\begin{align}
\Delta a_s(\tau, \lambda_1, \lambda_2) =
-&\frac{81}{4}
\,\tau^{2} (\tau-1)^{2} \times\nonumber \\
&\left[
3(\lambda_1-\lambda_2)\tau + (\lambda_2-2\lambda_1)
\right],
\end{align}
which has zero first derivatives (and is zero) at the boundary times.
The numerical optimization is then carried out over
\begin{align}
a_s^{(vn)}(t)  = a_s^{(v)}(t) + \Delta a_s(\tau, \lambda_1, \lambda_2),
\end{align}
in the superfluid regime and
\begin{align}
a_s^{(A)}(t)  = a_s^{(2)}(t) + \Delta a_s(\tau, \lambda_1, \lambda_2),
\nonumber \\
a_s^{(B)}(t)  = a_s^{(3)}(t) + \Delta a_s(\tau, \lambda_1, \lambda_2),
\end{align}
for the superfluid to supersolid transition.
A diagram of $\Delta a_s$ is shown in Fig. \ref{fig:app} (b).

To optimize $a_s^{(vn)}(t)$, $a_s^{(A)}(t)$ and $a_s^{(B)}(t)$ over $\lambda_1,\lambda_2$, the NLopt library was used which offers many efficient gradient and non-gradient search methods with a convenient C language API \cite{2007-NLoptNonlinearoptimizationPackage-Johnson-Johnson}.
For the results presented here, the DIRECT-L and Bound Optimization BY Quadratic Approximation (BOBYQA) algorithms were used, and both gave very similar results \cite{1993-LipschitzianOptimizationLipschitz-Jones-Stuckman,2001-LocallyBiasedFormDIRECT-Gablonsky-Kelley,2010-BOBYQAAlgorithmBound-Powell-Powell}.
In Fig. \ref{fig:sf-ss-fid}, the solid blue line in each plot comes from optimizing $a_s^{(A)}(t)$ and $a_s^{(B)}(t)$ using both algorithms, and both converged to the result shown that we label $a_s^{(n)}(t)$.
The DIRECT-L algorithm is useful for methodically finding a local minimum/maximum in a bounded region, but at the cost of many evaluations.
The BOBYQA algorithm in general requires fewer evaluations, using a quadratic approximation to search the parameter space, but needs a better chosen initial search point.
A combination of both methods is an efficient compromise, where the BOBYQA algorithm can be used to refine a relatively coarser DIRECT-L search.

\section{Implementation of the dipole interaction}\label{app:dip-term}
The Fourier transform of the dipole interaction is \cite{2022-DipolarPhysicsReview-Chomaz-Pfau}
\begin{align}\label{eq:U_tilde_sph}
\widetilde{U}_{dd} \left( \ve{k}\right) &= 
C_{dd} \left[ \cos^{2} \alpha - \frac{1}{3} \right],
\end{align}
where $\cos^{2} \alpha = k_{z}^{2} / k^{2} $.
While $\widetilde{U}_{dd}$ offers a much simpler numerical implementation in Fourier space than $U_{dd}$ in coordinate space, it requires a large numerical grid to be accurate in a finite region \cite{2006-BogoliubovModesDipolar-Ronen-Bohn, 2010-SpatialDensityOscillations-Lu-Yi} and there is also a discontinuity at $\ve{k} = 0$, as a result of the long-range and non-isotropic nature of the interaction \cite{2006-BogoliubovModesDipolar-Ronen-Bohn, 2010-SpatialDensityOscillations-Lu-Yi}.
Here we consider condensates that are finite and well localized, i.e. for some $R>0$, the condensate density $\fabsq{\Psi(\ve{r},t)}$ is negligible for $r>R.$
As is standard practice, this motivates the consideration of a spherically-truncated dipolar interaction \cite{2022-DipolarPhysicsReview-Chomaz-Pfau}
\begin{align}
U_{dd}^R(\ve{r}) = 
\begin{cases}
C_{dd}\left[1 - 3 \cos^{2}\theta\right]/r^3,  & r \leq R \\ \\
0, r>R,
\end{cases}
\end{align}
whose Fourier transform is then \cite{2022-DipolarPhysicsReview-Chomaz-Pfau}
\begin{align}\label{app:simple_trunc}
&\widetilde{U}_{dd}^R \left( \ve{k}\right) = 
C_{dd} \left[ 1 - 3 \cos^{2} \alpha\right]
\nonumber \\
& \hspace{0.87cm} \times 
\int_{0}^{ k R} du \:
\left[
\frac{\sin  u }{u^{2}} + 
\frac{3 \cos  u }{u^3} -
\frac{3 \sin  u }{u^4}
\right]
\nonumber \\
& = C_{dd} \left[ \frac{k_{z}^{2}}{k^{2}} - \frac{1}{3}\right]
\left[
1 + 
\frac{3 \cos\left( kR \right)}{(kR)^{2}} -
\frac{3 \sin\left( kR \right)}{(kR)^3}
\right].
\end{align}
$\widetilde{U}_{dd}^R$ is continuous at $\ve{k} = 0$, and is naturally suited for a numerical split operator implementation, since the Fourier space interaction can be applied efficiently using fast Fourier transforms \cite{2006-BogoliubovModesDipolar-Ronen-Bohn, 2010-SpatialDensityOscillations-Lu-Yi}.

However, for $\widetilde{U}_{dd}^R$ to be an accurate approximation of $\widetilde{U}_{dd}$, the condensate should be confined and simulated within a sphere of radius $R$.
This is not efficient for cigar or pancake-shaped traps that lead to the condensate being localized in cylindrical geometries.
This issue has been addressed in a number of ways in the literature; truncation of $U_{dd}$ to a cylindrical region \cite{2010-SpatialDensityOscillations-Lu-Yi} (but requires numerical evaluation of the Fourier transform integral over part of the integral domain), use of a non-uniform Fourier transform \cite{2014-FastAccurateEvaluation-Jiang-Bao}, use of a Gaussian sum approximation to the interaction \cite{2018-AnisotropicTruncatedKernel-Greengard-Zhang}, optimal padding of the interaction in Fourier space \cite{2024-OptimalZeroPaddingKernel-Liu-Zhang}, and recently a fast convolution based on a far-field smooth approximation of the Fourier transform of $U_{dd}$ \cite{2025-FastConvolutionSolver-Liu-Zhang,2025-EfficientHighOrderCompact-Liu-Zhang} (inspired by Ewald summation).
For simplicity we used the padding scheme from \cite{2024-OptimalZeroPaddingKernel-Liu-Zhang} to test whether the simple spherical truncation in \Eqref{app:simple_trunc} was accurate enough and found that the results obtained were not altered significantly.
Given that the eGPE is itself a mean-field approximation to the true dynamics of the system, a simple spherical truncation was used.


\bibliography{paper}

\end{document}